\documentclass[%
amsmath,amssymb,
aps,
pra,
floatfix,
twocolumn
]{revtex4-1}

\usepackage{graphicx}
\usepackage{dcolumn}
\usepackage{bm}
\usepackage[colorlinks=true, allcolors=blue]{hyperref}
\usepackage[nolist,nohyperlinks]{acronym}
\usepackage{multirow}
\usepackage[percent]{overpic}

\usepackage{braket}

\newacro{qec}[QEC]{quantum error correction}
\newacro{arma}[ARMA]{autoregressive-moving-average}

\begin{document}

\title{Impact of correlations and heavy-tails on quantum error correction}
\author{B.D. Clader}
\email{dave.clader@jhuapl.edu}
\author{Colin J. Trout}
\author{Jeff P. Barnes}
\author{Kevin Schultz}
\author{Gregory Quiroz}
\author{Paraj Titum}
\affiliation{Johns Hopkins University Applied Physics Laboratory\\
11100 Johns Hopkins Road, Laurel, MD 20723, USA}

\begin{abstract}
We show that space- and time-correlated single-qubit rotation errors can lead to high-weight errors in a quantum circuit when the rotation angles are drawn from heavy-tailed distributions. This leads to a breakdown of \acl{qec}, yielding reduced or in some cases no protection of the encoded logical qubits. While heavy-tailed phenomena are prevalent in the natural world, there is very little research as to whether noise with these statistics exist in current quantum processing devices. Furthermore, it is an open problem to develop tomographic or noise spectroscopy protocols that could test for the existence of noise with such statistics. These results suggest the need for quantum characterization methods that can reliably detect or reject the presence of such errors together with continued first-principles studies of the origins of space- and time-correlated noise in quantum processors. If such noise does exist, physical or control-based mitigation protocols must be developed to mitigate this noise as it would severely hinder the performance of fault-tolerant quantum computers. 
\end{abstract}

\maketitle

\section{Introduction}\label{sec:intro}
The theory of fault tolerance and the associated threshold theorem demonstrate that if the physical error rate per gate can be lowered below some threshold, then one can perform quantum computation with arbitrary accuracy with polynomial overhead (see e.g. \cite{shor1995, steane1996, aliferis2006, aharonov2008, terhal2015}). The prevailing noise model for analyzing quantum error correcting codes is noise that manifests itself as bit flips and phase flips that are local in both time and space, meaning there are no spatial or temporal correlations \cite{nielsen2010}. However, it is well known that at least some spatial correlations are inevitable in a quantum system due to such physical effects as common baths shared amongst qubits or control-line crosstalk \cite{harper2020}. In addition, time-correlated noise is generally always present (e.g. $1/f^\alpha$ type noise in superconducting qubit system \cite{PhysRevLett.99.187006, Bylander:2011wc, Burnett:2014tv, RevModPhys.86.361, PhysRevApplied.6.041001, Burnett:2019ti}).

For these reasons, there have been a number of studies that have examined the impact of spatial and temporal noise correlations on quantum error correction (see e.g. \cite{clemens2004, klesse2005, terhal2005, aharonov2006, aliferis2006, alicki2006, novais2006, novais2008, shabani2008}). Despite the proliferation of studies examining noise correlations, the complete impact on error correction is mixed. Some studies suggest that noise correlations are fairly detrimental \cite{clemens2004, klesse2005, alicki2006, novais2008, cafaro2010} while others suggest that most realistic models of correlated noise can still be handled via quantum error correction with a manageable overhead \cite{clemens2004, terhal2005, aharonov2006, aliferis2006, novais2006, shabani2008}.

These disagreements arise due to the difficulty in analyzing and simulating a \acf{qec} code operating on realistic multi-level quantum systems interacting with a general open quantum system bath. Approximations are always required to create manageable calculations such as considering only the two-level subspace of the multi-level system and severe approximations as to the impact of the bath on the qubits. The most restrictive of these bath approximations is the assumption that it can be modeled using the Pauli error model where random bit flips and phase flips are inserted stochastically into the circuit. Despite the simple nature of this error model, it has proven highly useful in the general theory of fault-tolerant \ac{qec} and the development of new \ac{qec} codes. 

The standard theory of fault-tolerant \ac{qec} assumes that if a weight-one Pauli error occurs with probability $p$ then a weight-two Pauli error occurs with probability of order $p^2$ and so on. Under these assumptions, a \ac{qec} code will yield a logical error rate proportional to $p^{(d+1)/2}$ where $d$ is the code distance. Correlated Pauli errors are clearly detrimental and reduce the effectiveness of \ac{qec}. If a weight-two error occurs with an error rate proportional to $p$ then the effective distance of the code is reduced by 1 and so on. Generating these types of errors can happen if the interaction Hamiltonian contains entangling terms (sometimes called weight-two generators). These types of errors might arise from a common bath shared by all qubits that generates entanglement or residual coupling between qubits that cannot be effectively turned off to a sufficient level. It is generally assumed that qubits at best might share some portion of the bath, but that the correlations would be short range at worst. Unfortunately, models of open quantum systems with spatially correlated baths are not well studied, although some progress has been made \cite{jeske2013}. Despite this, it is generally assumed that these higher weight errors can be effectively controlled and mitigated in future quantum systems.

In this report, we analyze a related model of decoherence, but one that avoids the difficulties required when performing general open quantum system calculations. The errors are modeled as classical random variables within the Hamiltonian. This model is often called the semi-classical system bath model. It is an alternative model to a general open quantum system coupled to a quantum bath, and is considered a reasonable approximation in certain instances. In particular, the bath must be in thermal equilibrium, there is no back action on the environment from the qubits (i.e. the bath dephases instantly), and the bath is at infinite temperature yielding equal populations of the qubit states after long term decay \cite{Kubo1963, Haken1973, Capek1993, Cheng2004, gardiner2004, Cheng2005, vankampen2007}. These assumptions apply directly to classical noise from the classical control system. The advantage to such a model from our perspective is that spatial and temporal correlations are fairly straightforward to model in this manner. There are known situations where this approximation breaks down and the quantum nature of the bath leads to observable signatures such as nontrivial phase evolution in addition to pure decoherence \cite{Paz-Silva2017}. This can occur for example for baths at low temperature or systems that are strongly coupled. We do not consider these more general quantum bath models here, and leave that for future work.

Using the semiclassical noise model allows us to demonstrate a rather surprising result. We show that single-qubit rotation errors arising from weight-one error generators can lead to high-weight errors in a quantum system when the noise is either spatially or temporally correlated and drawn from certain types of heavy-tailed distributions \cite{bryson1974}. This results in a reduction in the effective distance of the \ac{qec} code that depends on the tails of the distribution. Various definitions of heavy-tailed (sometimes referred to as fat-tailed) distributions exist. Generally speaking they are distributions whose tails decay more slowly than exponential (e.g. as a power law) and they have undefined (or infinite) variance. They are well-studied in the quantitative finance literature \cite{Haas2009} as they are routinely used to model things such as exogenous shocks to financial markets (see e.g. Refs. \cite{Villaverde2007, Fagiolo2008, Mishkin2011, ascari2012}) or even the value of returns for asset prices recognized early on with the seminal work of Mandelbrot \cite{Mandelbrot1963}.

Relatively little research has been conducted as to whether semi-classical noise with heavy-tailed distributions exists in quantum computing devices, but this not true for quantum systems in general. Events with heavy-tailed distributions have been discussed extensively in the context of physical models that generate $1/f^\alpha$ noise spectra.  It has been shown that signals from systems with dynamics that adhere to families of point process models \cite{Niemann2013,Eliazar2010,Davidsen2002,Kaulakys2005,Lowen1993,Lukovic2008} and linear/nonlinear stochastic differential equations (SDEs) \cite{Kaulakys2013,Ruseckas2010} generate $1/f^\alpha$ spectra when the signal probability density functions (PDFs) obey heavy-tailed statistics.  Various physical systems have exhibited signals with power-law statistics.  Similar point process models have been utilized to describe the power-law behavior of fluorescent blinking in quantum dots \cite{Kuno2000,Shimizu2001,Pelton2004,Margolin2006,Pelton2007,Mahler2008,Frantsuzov2009,Frantsuzov2013} and single-molecule fluorescence of organic molecules \cite{Haase2004,Schuster2005,Hoogenboom2005,Yeow2006}.  Power-law behavior has been observed in trapping times for charge transport in amorphous semiconductors \cite{Lowen1993,Dhariwal1991,Tiedje1981,Lowen1992,PhysPropAmorphMat} and nanoscale electrodes \cite{Krapf2013}.  The family of SDEs that exhibit power law behavior generate dynamics that violate the fluctuation dissipation and equipartition of energy theorems which has been observed in finite-dimensional spin-glasses \cite{Lobaskin2006}.  Furthermore, spin glasses exhibit random couplings and relaxation rates that obey power-law behavior  \cite{Klein1968, Klein1976, Cizeau1993, Berkov1996, Pickup2009, Neri2010}.

For quantum processors, non-Gaussian noise spectroscopy techniques have been developed \cite{Norris2016} and demonstrated experimentally \cite{Sung2019}, but this approach only applies to noise with distributions tighter than Gaussian. The difficulty in developing characterization techniques for heavy-tailed distributions is that most quantum characterization techniques that seek to characterize the statistics of noise correlations rely on expanding the statistics of the noise into moments (or cumulants) \cite{Kubo1962, VanKampen1974-215, VanKampen1974-239}. Since higher-order moments are undefined or infinite for heavy-tailed distributions these existing techniques do not apply. Our results show that a new characterization technique is needed to test for this type of harmful noise in quantum systems. If this type of noise is present, it is imperative that physical or control-based mitigation schemes are developed to reduce its effect as \ac{qec} will not suffice.

\section{Impact of correlated noise on QEC: Analytic Model}

We begin our analysis by considering an analytic model of \ac{qec} where we consider noise arising from stochastic unitary rotation errors on the data qubits only in an $n$-qubit perfect code. This model is referred to as the code-capacity error model. Physically, these correlations would correspond to spatial correlations between qubits. The term perfect here refers to the fact that our analytical analysis assumes that a distance $d$ code can correct exactly $(d-1)/2$ errors on the encoded qubits and no more. Certain codes, such as surface codes, can correct certain types of high-weight errors that will break this assumption. Numerical simulations shown later demonstrate that our code-capacity model accurately predicts the behavior of fully fault-tolerant implementations of \ac{qec} with noise affecting data and ancilla qubits at all locations in the circuit.

Our model starts by assuming that we have perfectly encoded an $n$-qubit logical state into a \ac{qec} code denoted as $\ket{\psi_L}$. We then apply a single-qubit rotation about an arbitrary axis to all of the data qubits in the encoded state yielding
\begin{equation}
\label{eq:Nqubitlogicalstate}
\ket{\psi_L} \to \prod_{j=1}^n\left[ \cos\left(\frac{\theta_j}{2}\right)\hat{I} - i \sin \left(\frac{\theta_j}{2}\right)\vec{v}\cdot\hat{\vec{\sigma}}^{(j)}\right] \ket{\psi_L},
\end{equation}
where $\hat{I}$ is the identity matrix, $\vec{v}$ is an arbitrary unit three vector of real numbers specifying the direction of rotation, $\hat{\vec{\sigma}}^{(j)}$ is a three vector containing the Pauli $x, y$, and $z$ operators, and $\theta_j$ is the angle of rotation for qubit $j$. We consider two cases. In the uncorrelated case, each angle of rotation is drawn from a probability distribution and taken to be independent for a total of $n$ independent random variables. In the correlated case, we assume that a single angle is drawn from the same probability distribution and applied to each qubit. The average probability of a logical error is then given by
\begin{subequations}
\label{eq:nqubit}
\begin{equation}
P_{unc} =1 - \sum_{k=0}^w \binom{n}{k} \left\langle \cos^2 \left(\frac{\theta}{2}\right)\right\rangle^{n-k}\left\langle\sin^2 \left(\frac{\theta}{2}\right)\right\rangle^k
\end{equation}
\begin{equation}
P_{cor} = 1- \sum_{k=0}^w \binom{n}{k} \left\langle \cos^{2(n-k)} \left(\frac{\theta}{2}\right) \sin^{2k} \left(\frac{\theta}{2}\right)\right\rangle,
\end{equation}
\end{subequations}
where we have assumed that our error correcting code can correct all Pauli errors of weight $w=(d-1)/2$, $\theta$ is a random variable, and the angular brackets $\langle \cdot \rangle$ denote an ensemble average. After some algebraic manipulations, we can transform Eqs. \eqref{eq:nqubit} into 
\begin{widetext}
\begin{subequations}
\label{eq:nqubit_cf}
\begin{equation}
P_{unc} = 1 - \frac{1}{2^n} \sum_{k=0}^w \sum_{l=0}^{n-k}\sum_{m=0}^k\binom{n}{k}\binom{n-k}{l}\binom{k}{m}(-1)^{m-k}f(t=1)^{n-l-m}
\end{equation}
\begin{equation}
P_{cor} = 1 - \frac{1}{2^{2n}} \sum_{k=0}^w \sum_{l=0}^{2(n-k)}\sum_{m=0}^{2k}\binom{n}{k}\binom{2(n-k)}{l}\binom{2k}{m}(-1)^{m-k}f(t=n-l-m),
\end{equation}
\end{subequations}
\end{widetext}
where $f(t)$ is the characteristic function of the probability distribution. In deriving Eqns. \eqref{eq:nqubit_cf} we have assumed distributions that are symmetric about 0 such that the characteristic function is even $f(t) = f(-t)$. For a single physical qubit, we define the failure probability to be the probability that upon measurement we get either a bit-flip or phase-flip error. For a qubit rotated by an angle $\theta$, this is given by
\begin{equation}
\label{eq:physical_error_rate}
P = \sin^2\left(\frac{\theta}{2}\right)
\end{equation}
with the corresponding expectation value in terms of the characteristic function given by
\begin{equation}
\label{eq:physical_error_rate_expectation}
\langle P \rangle \equiv P_{ph} = \frac{1}{2}\left[1-f(t=1)\right],
\end{equation}
where $P_{ph}$ stands for the probability of a physical error occurring. Evaluation and comparison of Eqs. \eqref{eq:nqubit_cf} and \eqref{eq:physical_error_rate_expectation} allow us to examine the impact of various noise distributions by inserting the known characteristic function and computing the formula for given code sizes and comparing the logical error rate with the physical error rate. For the purpose of our analysis, the number of qubits in the code will be equal to the the Knill--Laflamme bound $n \ge 4w+1$ unless otherwise specified.

\begin{table*}
\caption{\label{tab:student_results} Comparison of the leading order term in a $\sigma \ll 1$ expansion of the failure probability for a physical qubit and various distance perfect codes with uncorrelated and correlate noise. The rows are for various $\nu=2r-1$ parameters in the Student's $t$-distribution as given in Eq. \eqref{eq:students_pdf}. As $r$ increases the tails of the distribution are reduced, resulting in a tighter distribution. As this occurs, the reduction in effective code distance is pushed out to higher distances. As an example, for a $d=3$ code at $r=3$ the effective distances are equivalent for uncorrelated and correlated noise. We find that when $d \le 2r-3$ the code distance is equal for correlated and uncorrelated noise, but when $d>2r-3$ the effective distance is reduced for correlated noise. For all code distances we set the number of qubits to $n=4w+1$, which is the minimum Knill-Laflamme bound.}
\begin{ruledtabular}
\begin{tabular}{r|r|rr|rr|rr|rr}
$r$ &\hspace{-0.1in}Physical & \multicolumn{2}{c|}{$d=3$}                                        &  \multicolumn{2}{c|}{$d=5$}                                      & \multicolumn{2}{c|}{$d=7$}                                        & \multicolumn{2}{c}{$d=9$}  \\
  {}  &    {}                                & \multicolumn{1}{c}{Unc} & \multicolumn{1}{c|}{Cor}  & \multicolumn{1}{c}{Unc} & \multicolumn{1}{c|}{Cor}  & \multicolumn{1}{c}{Unc} & \multicolumn{1}{c|}{Cor}  & \multicolumn{1}{c}{Unc} & \multicolumn{1}{c}{Cor} \\
\hline
\\[-2ex]
1 & $\frac{1}{2}\sigma^{\phantom{1}}$ &          $\frac{5}{2}\sigma^2$ & $\frac{35}{64}\sigma^{\phantom{1}}$ &                     $\frac{21}{2}\sigma^3$ & $\frac{9009}{16384}\sigma^{\phantom{1}}$ &                          $\frac{715}{16}\sigma^4$ & $\frac{1154725}{2097152}\sigma^{\phantom{1}}$ &                        $\frac{1547}{8}\sigma^{5\phantom{1}}$ & $\frac{591534125}{1073741824}\sigma^{\phantom{10}}$  \\[1ex] \hline
\\[-2ex]
2 & $\frac{3}{4}\sigma^2$ &      $\frac{45}{8}\sigma^4$ & $\frac{175\sqrt{3}}{64}\sigma^3$ &  $\frac{567}{16}\sigma^6$ & $\frac{27027\sqrt{3}}{16384}\sigma^3$ &     $\frac{57915}{256}\sigma^8$ & $\frac{3002285\sqrt{3}}{2097152}\sigma^3$ &   $\frac{375921}{256}\sigma^{10}$ & $\frac{1436582875\sqrt{3}}{1073741824}\sigma^{3\phantom{1}}$  \\[1ex] \hline 
\\[-2ex]
3 & $\frac{5}{12}\sigma^2$ &    $\frac{125}{72}\sigma^4$ & $\frac{125}{8}\sigma^4$ &             $\frac{875}{144}\sigma^6$ & $\frac{705705\sqrt{5}}{16384}\sigma^5$ &   $\frac{446875}{20736}\sigma^8$ & $\frac{345262775\sqrt{5}}{18874368}\sigma^5$ &      $\frac{4834375}{62208}\sigma^{10}$ & $\frac{130729041625\sqrt{5}}{9663676416}\sigma^{5\phantom{1}}$  \\[1ex] \hline 
\\[-2ex]
4 & $\frac{7}{20}\sigma^2$ &    $\frac{49}{40}\sigma^4$ & $\frac{49}{8}\sigma^4$ &                $\frac{7203}{2000}\sigma^6$ & $\frac{7203}{16}\sigma^6$ &                     $\frac{343343}{32000}\sigma^8$ & $\frac{113011411513\sqrt{7}}{94371840}\sigma^7$ &   $\frac{26000429}{800000}\sigma^{10}$ & $\frac{3787119775075\sqrt{7}}{9663676416}\sigma^{7\phantom{1}}$  \\[1ex] \hline 
\\[-2ex]
5 & $\frac{9}{28}\sigma^2$ &    $\frac{405}{392}\sigma^4$ & $\frac{243}{56}\sigma^4$ &       $\frac{2187}{784}\sigma^6$ & $\frac{2187}{16}\sigma^6$ &                    $\frac{4691115}{614656}\sigma^8$ & $\frac{4691115 }{256}\sigma^8$ &                     $\frac{13049829}{614656}\sigma^{10}$ & $\frac{1086610719708657}{7516192768}\sigma^{9\phantom{1}}$  \\[1ex] \hline 
\\[-2ex]
6 & $\frac{11}{36}\sigma^2$ &   $\frac{605}{648}\sigma^4$ & $\frac{605}{168}\sigma^4$ &     $\frac{9317}{3888}\sigma^6$ & $\frac{1331}{16}\sigma^6$ &               $\frac{10468315}{1679616}\sigma^8$ & $\frac{10468315}{2304}\sigma^8$ &            $\frac{249145897}{15116544}\sigma^{10}$ & $\frac{249145897}{256}\sigma^{10}$  \\[1ex]
\end{tabular}
\end{ruledtabular}
\end{table*}

\begin{figure}
\centering
\includegraphics[width=\columnwidth]{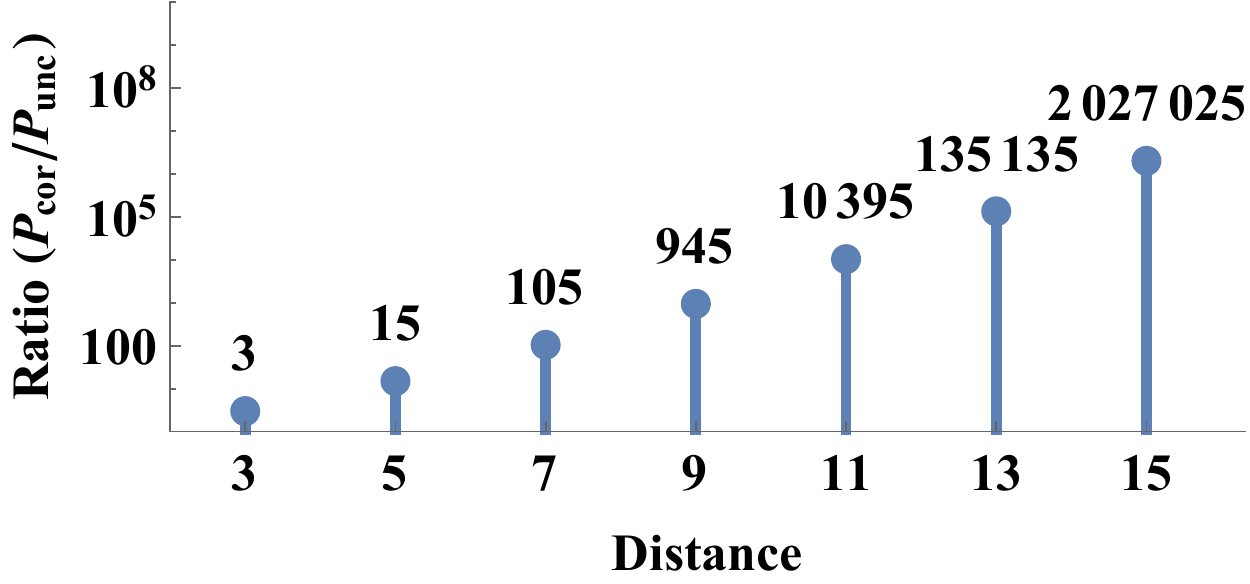}
\caption{\label{fig:ratio}Ratio of $P_{cor}/P_{unc}$ of the leading order term in powers of $\sigma$ in the error expansion of Eqs. \eqref{eq:nqubit_cf} for correlated and uncorrelated Gaussian noise, assuming $\sigma \ll 1$. The labels on the data points correspond to the numerical value of the ratio. As the code distance increases, the ratio increases as $d!!$.}
\end{figure}

\subsection{Gaussian}
First, we examine the case where the noise is drawn from a Gaussian distribution with zero mean and standard deviation $\sigma$. The characteristic function of a Gaussian random variable with zero mean is
\begin{equation}
\label{eq:gauss_cf}
f(t; \sigma) = e^{-\frac{1}{2}\sigma^2 t^2}.
\end{equation}
To understand the impact on \ac{qec}, we study the low noise scaling of the physical and logical error rates with respect to the width parameter of the probability distribution. For the Gaussian distribution this implies taking a series expansion about $\sigma \to 0$ for Eqs. \eqref{eq:nqubit_cf}. Evaluation of the leading order terms for both correlated and uncorrelated noise reveals logical error rates that scale as $\sigma^{2(w+1)}$ where $w$ is the number of correctable errors for a code of distance $d=2w+1$. Thus, Gaussian correlated noise does not yield high-weight errors that might impact the code distance, however it does affect the series coefficient. To see this, we plot the ratio of the first non-zero term for correlated noise to uncorrelated noise in Fig. \ref{fig:ratio}. As the distance of the code increases, the ratio of the series coefficient grows at a rate proportional to $d!!$ implying that the logical error rate is increased by a related factor for correlated noise. This implies that if correlated single-qubit rotation noise drawn from a Gaussian distribution is present, its impact may be minor for small codes, but grows as code distance increases. We note as an aside that the $\sigma \to 0$ approximation of Eqs. \eqref{eq:nqubit} must be done with care as the number of terms in the binomial expansion grows for larger codes. Thus the approximation must technically be in the regime where $\sigma \ll (1/d!!)^{2/(d+1)}$. These results agree with previously published work and imply that a threshold does not exist \cite{clemens2004}. Despite this, the code can still suppress noise and a pseudo-threshold does exist for each distance.

For a code of distance 3, the leading order terms assuming $\sigma \ll 1$ are
\begin{align}
\label{eq:gauss_expvalues}
P_{ph} & \approx \frac{\sigma^2}{4} \\ \nonumber
P_{unc} & \approx \frac{5 \sigma^4}{8} \\ \nonumber
P_{cor} &  \approx \frac{15 \sigma^4}{8}.
\end{align}
The logical error rate for both correlated and uncorrelated noise is reduced by a squared factor of $\sigma$, while the pre-factor of the correlated noise logical error rate increases by a factor of three relative to the uncorrelated case. The quadratic reduction of the logical error rate suggests the code is behaving as expected by a distance $3$ code in both cases.

\subsection{Student's t}
Next, we examine the Student's $t$-distribution to see what impact varying the probability distribution of the random variable has on the physical and logical error rates. The Student's $t$-distribution is an example of a heavy-tailed distribution with a discrete parameter $\nu$ that governs the heaviness of the tails of the distribution. This can be seen by considering the probability density function of the Student's $t$-distribution
\begin{equation}
\label{eq:students_pdf}
PDF_{t}(\theta; \nu, \sigma) = \frac{(\nu\sigma^2)^{\frac{\nu}{2}}\Gamma\left(\frac{\nu + 1}{2}\right)}{\sqrt{\pi}\Gamma\left(\frac{\nu}{2}\right)}\frac{1}{\left(\nu\sigma^2+\theta^2\right)^{\frac{\nu+1}{2}}}.
\end{equation}
Here $\nu\ge1$ is an integer corresponding to the number of degrees of freedom of the distribution and $\Gamma(x)$ is the gamma function. We will restrict our attention to cases where $\nu$ is odd. Taking $\nu=1$ gives exactly the Cauchy distribution, while larger values of $\nu$ tightens the tails of the probability distribution via this discrete parameter. In the limit of $\nu \to \infty$ the Gaussian distribution is recovered. The characteristic function of a Student's $t$-distribution is
\begin{equation}
\label{eq:students_cf}
f(t;\sigma,\nu) = \frac{\sigma^{\frac{\nu}{2}}\nu^{\frac{\nu}{4}}|t|^\frac{\nu}{2}}{2^{\frac{\nu}{2}-1}\Gamma\left(\frac{\nu}{2}\right)}K_{\nu/2}\left(\sigma\sqrt{\nu}|t|\right),
\end{equation}
where $K_n(x)$ is the modified Bessel function of the second kind.

Again, we just report the approximate expressions in the $\sigma \ll 1$ limit. For additional ease of analysis, we take $\nu$ to be odd, but we lift this restriction in our numerical simulations. We display the results in Table \ref{tab:student_results} for the physical and logical errors rates with both correlated and uncorrelated noise, where we have replaced $\nu$ with $\nu=2r-1$ and $r\ge 1$ is a positive integer. For correlated Cauchy random noise ($r=1$) error correction provides no error suppression as the effective code distance is reduced to 1 for all code distances considered. We conjecture that this extends to arbitrary code distance. This implies that single-qubit rotations with correlated rotation angles drawn from a Cauchy distribution result in at least weight $(d+1)/2$ errors in the quantum circuit. As $r$ increases the tails of the distribution are tightened, and the reduction in the code distance is pushed out to higher distances. We find that when the distance is $d \le 2r-3$, for positive $d$, the effective distance of the \ac{qec} is equivalent for the correlated and uncorrelated cases. However, once $d > 2r-3$ the effective distance begins to be reduced for correlated noise with the maximum exponent on $\sigma$ appearing to be $2r-1$ for correlated noise. These results imply that correlated single-qubit noise drawn from the Student's $t$-distribution yields higher-weight Pauli errors with the weight of the error related to the tail index of the distribution. 

\begin{figure}
\centering
\includegraphics[width=\columnwidth]{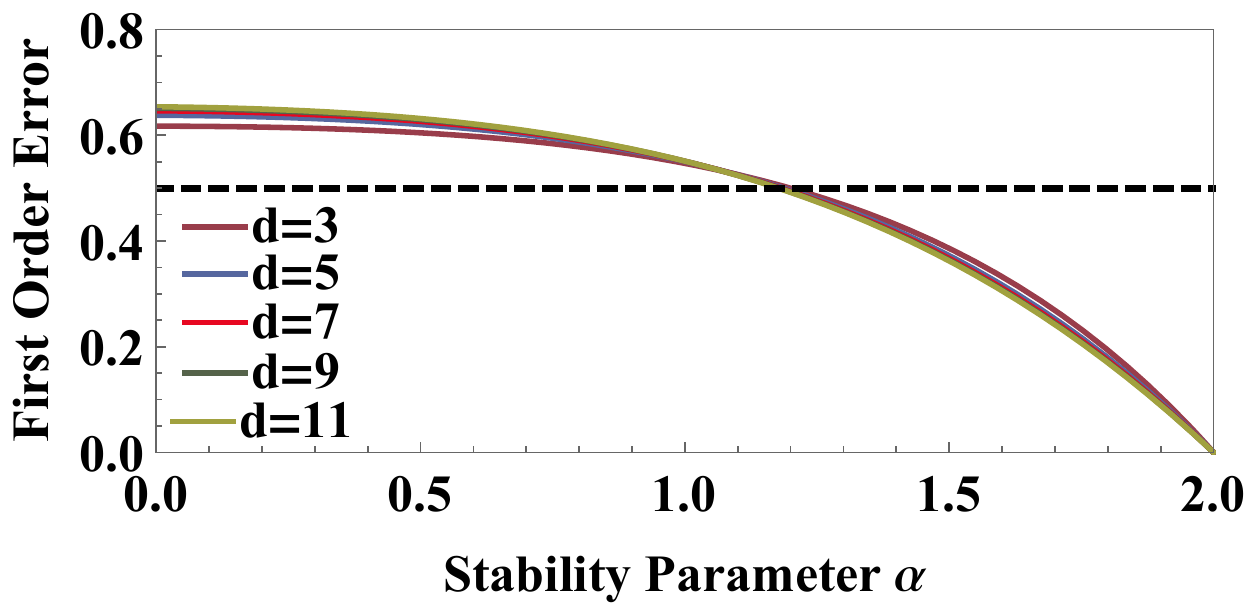}
\caption{\label{fig:stable_analytical}The various plots correspond to the value of the error term proportional to $\sigma^{\alpha}$ for all valid values of $\alpha$ in the correlated noise error expansion, $P_{cor}$. The dashed line is the first order term of the physical error rate $P_{ph} =0.5$ of Eqs. \eqref{eq:nqubit_cf}. Code distance does not have a large impact as seen by the nearly indistinguishable curves. For all cases plotted, the logical error rate containing correlated noise from an alpha stable distribution will not provide protection of the encoded logical qubits as this distribution results in errors with weight of $(d+1)/2$ or greater expect at the exact point where $\alpha=2$, which corresponds to the Gaussian distribution.}
\end{figure}

\subsection{L\'evy alpha-stable}
Finally, we consider the L\'evy alpha-stable distributions. These are a family of probability distributions that contain the Gaussian and Cauchy distributions and allow for continuous interpolation between them. It is also useful as it is a stable distribution meaning that sums of independent random variates have the same distribution up to location and scale parameters. This allows us to efficiently model time correlations in our numerical simulations that we show in the next section. The probability density function for the stable distribution is not analytically expressible. However the characteristic function is expressible. We write a simplified form of the characteristic function where we take the skewness and location parameters to be zero yielding
\begin{equation}
\label{eq:stable_cf}
f(t;\sigma,\alpha) = e^{-|\sigma t|^\alpha}.
\end{equation} 
The parameter $\alpha$ is  the stability parameter and it lies in the range $(0,2]$. The stable distribution corresponds to a Gaussian when $\alpha=2$ and a Cauchy when $\alpha=1$, and allows for continuous interpolation between Gaussian and heavy-tailed distributions. The formulas are complicated for the general case, so we examine the simplest case for a distance 3 code. The leading order terms, assuming $\sigma \ll 1$, for the physical and logical error rates are
\begin{align}
\label{eq:stable_expvalues}
P_{ph} &  \approx \frac{\sigma^\alpha}{2} \\ \nonumber
P_{unc} & \approx \frac{5 \sigma^{2\alpha}}{2} \\ \nonumber
P_{cor} & \approx \frac{1}{128} \left[5 \left(2^{\alpha +2}\right)-5\left(3^{\alpha}\right)-5\left(4^{\alpha}\right)-5^{\alpha}+70\right]\sigma^{\alpha} \\ \nonumber
 & +\frac{1}{256} \left[5 \left(-4^{\alpha +1}+9^{\alpha }+16^{\alpha }-14\right)+25^{\alpha }\right]\sigma^{2\alpha}.
\end{align}
Only for exactly $\alpha=2$, which corresponds to a Gaussian distribution, does the lowest order term (the term of order $\sigma^{\alpha}$) of the correlated error rate disappear. Any $\alpha<2$ yields a term that is proportional to the physical error rate, causing the error correcting code to become ineffective against this type of noise. To examine larger code distances, we plot the value of the term proportional to $\sigma^\alpha$ for the correlated noise logical error rate in Fig. \ref{fig:stable_analytical}. The figure shows that for all code distances considered the logical error rate scaling is proportional to the physical error rate scaling. In other words, the \ac{qec} code offers no protection for this type of noise no matter how large of a code is used. Only exactly when $\alpha=2$ does the term dissapear and we recover protection of the encoded logical state as shown previously in the discussion on Gaussian noise.

\section{Numerical Simulations}
Our analytic model has shown the detrimental effect of spatially correlated noise on the data qubits within a single error correction block, with correlated single-qubit (weight one) rotation errors leading to uncorrectable multi-qubit (high-weight) errors. However, determining fault tolerance also requires the accounting of errors within the time-length of a decoding block \cite{aliferis2006}. Therefore, we anticipate that time correlations of weight-one error generators would have an equally harmful impact on quantum error correction as they would lead to a breakdown of fault-tolerance due to high-weight errors occurring in time across the decoding boundaries. To study this, we use numerical simulations of low-distance surface codes to examine the case of time-correlated errors from heavy-tailed distributions. In addition to perfect correlations that we studied with our analytical model, we also generalize to the situation where the noise correlations are defined by a correlation function. Our simulations show that time-correlations do result in a similar breakdown as our analytical results predict for spatially-correlated noise on the data qubits only, but with non-trivial dependence on the correlation function of the noise.

Because our errors are stochastic unitary errors, we must simulate the entire state-vector. This limits us to low-distance codes and for this report, we limit ourselves to just simulations of a distance-3 rotated surface code. We make use of the same simulation framework and circuits that we used in our study of random coherent errors \cite{Barnes2017}. We model the errors as random unitary gates $U_k^{(\ell)}(\theta_k^{(\ell)}) = \exp(-i\theta_k^{(\ell)}\sigma_y^{(\ell)})$ that create a rotation by a random angle $\theta_k^{(\ell)}$ (here drawn from Gaussian, Cauchy, Student's t, and L\'{e}vy alpha-stable distributions) about the $Y$ axis at a circuit time location $k$ for qubit $\ell$. We insert these errors across all the qubits in the code (both data and ancilla) after every single gate in the circuit. To compute the performance of the code we start with a random initial state
\begin{equation}
\label{eq:random_initial_state}
\ket{\psi_0} = \cos \alpha \ket{0} + e^{i \beta} \sin \alpha \ket{1},
\end{equation}
where $0 \le \alpha < 2\pi$ and $0 \le \beta < 2\pi$ are both uniform random variables. This random initialization covers the Bloch sphere, but it is not uniform. We define the fidelity to be 
\begin{equation}
\label{eq:physical_fidelity}
\mathcal{F}^2 = \frac{1}{(2\pi)^2}\iint_{0}^{2\pi}d\alpha d\beta\int_{-\infty}^{\infty} d\theta p(\theta) |\braket{\psi_0 | e^{-i \theta \sigma_y} | \psi_0}|^2,
\end{equation}
where $p(\theta)$ is the probability distribution for the error terms. This expression simply computes the overlap of the initial state and the final state and averages over the probability distributions. We note, as before, that this is not the standard definition of fidelity, since the distribution of $\alpha$ and $\beta$ is uniform this is not a Haar average over the Bloch sphere. This yields slightly different error rates for different errors on different axes, but this does not have any meaningful impact on our overall results. Using Eqs. \eqref{eq:random_initial_state} and \eqref{eq:physical_fidelity}, it is straightforward to analytically calculate the physical fidelity for the various probability distributions considered in this manuscript. It is
\begin{equation}
\label{eq:physical_fidelity_exact}
\mathcal{F}^2 = \frac{5}{8}+\frac{3}{8}f(t=2)
\end{equation}
where $f(t=2)$ is the characteristic function of the probability distribution and is given in  Eqs. \eqref{eq:gauss_cf}, \eqref{eq:students_cf}, and \eqref{eq:stable_cf} for the Gaussian, Student's t, and Stable distributions respectively.

To calculate the logical fidelity, we perfectly encode the random state defined in Eq. \eqref{eq:random_initial_state}, simulate three rounds of faulty syndrome extraction with errors inserted after every location where a gate exists (a circuit-level noise model), and follow that by decoding and perfect correction. We conclude the simulation with a round of perfect error correction to correct any trailing errors. We estimate the logical fidelity by numerically estimating the integral given in Eq. \eqref{eq:physical_fidelity} by computing the overlap of the final decoded logical state with the initial state and Monte Carlo sampling over the initial random states and error terms. We use bootstrap resampling to report the $95\%$ confidence intervals with $10^3$ samples with replacement used. Each data point is the result of $10^7$ independent trials.

For arbitrary time correlations we leverage SchWARMA \cite{schultz2020schwarma} to simulate time-correlated noise in quantum circuits. The SchWARMA method leverages a classical time-series modeling approach called \ac{arma} models where the angle of rotation at circuit time location $k$ is an \ac{arma} model
\begin{equation}\label{eq:ARMA}
   \theta_k^{(\ell)} = \underbrace{\sum_{i=1}^p a_{i}
    \theta_{k-i}^{(\ell)}}_{AR}+\underbrace{\sum_{j=0}^{q} b_{j}x_{k-j}^{(\ell)}}_{MA}\,.
\end{equation}
The set $\{a_i\}$ defines the autoregressive portion of the model, and $\{b_j\}$ the moving-average portion with $p$ and $q+1$ elements of each set, respectively, and the $x_{k-j}^{(\ell)}$ are random variables drawn from the user-defined probability distribution. Because \ac{arma} models require one to add random variates we must restrict ourselves to stable distributions when using this method to ensure that the probability distribution of the output model remains the same as the random variables. The Student's $t$-distribution is not stable. Therefore, in that case we consider white noise (uncorrelated in time) and direct-current (DC) noise in which we draw a single random variable at the beginning of the quantum circuit for each qubit and we use that same angle at all subsequent times in the circuit. These two limits correspond to an ARMA model with $p=0, q=0$ and $p=0, q\to\infty$ respectively. For the other cases we can consider more general time correlations. For the purposes of this study, we interpolate between white noise and DC noise by considering exponential moving-averages (EMAs) where the terms $b_j = N \exp(-\ln(2) j /T_h)$ where $N$ is chosen such that $\sum_j b_j = 1$ and $T_h$ is the ``half-life" of the moving average. We do not consider the AR portion in this paper, so $p=0$. We set each gate in the circuit to take a single unit of time, so the parameter $j$ corresponds exactly to the circuit depth to that point and we set the number of terms in the moving average to $q=10\lceil T_h\rceil$. The various syndrome extraction circuits for the surface code take anywhere from 2 to 6 time ticks in those units.

\subsection{Gaussian}
\begin{figure}[t]
\centering
\includegraphics[width=\columnwidth]{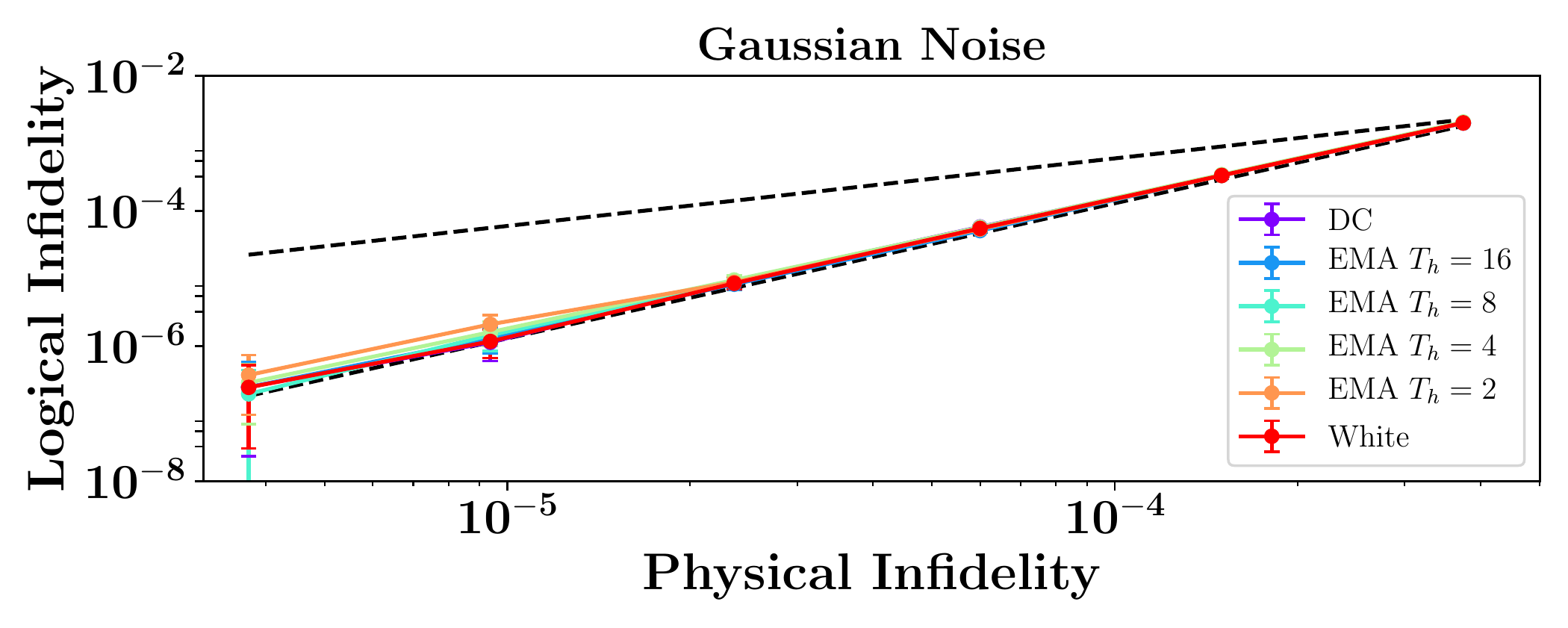}
\caption{\label{fig:gaussian_cheese}Pseudo-threshold plots obtained from simulating the rotated distance-3 surface code with independent and time-correlated single-qubit rotation errors drawn from a Gaussian distribution. The SchWARMA simulations denoted with titles EMA $T_h$ interpolate between the white and DC noise cases by using an exponential moving average filter. The various curves are denoted by their color, but are indistinguishable on this scale. This demonstrates that time-correlated noise drawn from a Gaussian distribution has minimal impact to low distance \ac{qec} codes. The top and bottom dashed lines show slopes where the logical infidelity is proportional to the physical infidelity and the square of the physical infidelity respectively. Error bars are the 95\% confidence intervals obtained from bootstrap resampling. They are cutoff at the lowest error rates for display purposes.}
\end{figure}

The results for Gaussian noise are plotted in Fig. \ref{fig:gaussian_cheese} where we plot the logical infidelity versus the physical infidelity. There is no discernible difference between the DC and white noise case. Below the pseudo-threshold the logical error rate is reduced relative to the physical error rate and the reduction is quadratic as expected for a distance three code. This is all in agreement with our analytical results given in Eq. \eqref{eq:gauss_expvalues}. 
\begin{figure}[h]
\centering
\includegraphics[width=\columnwidth]{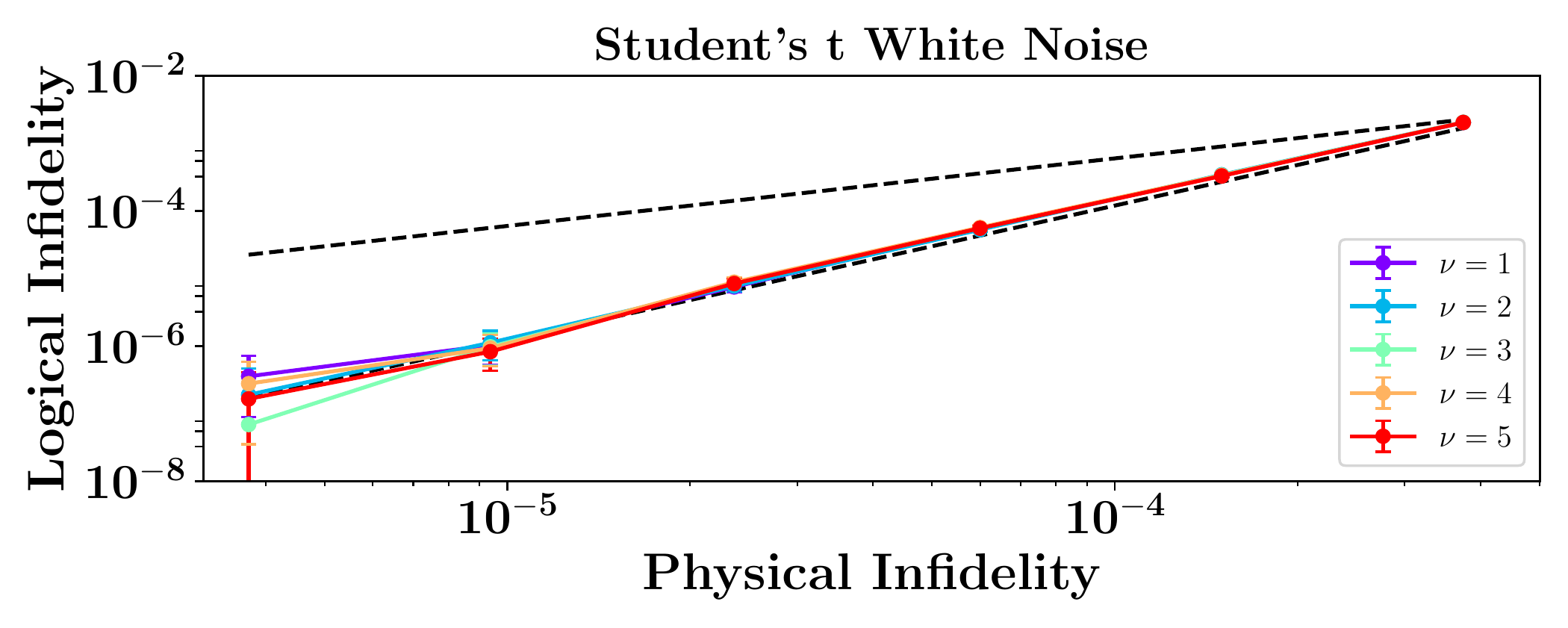}
\\
\includegraphics[width=\columnwidth]{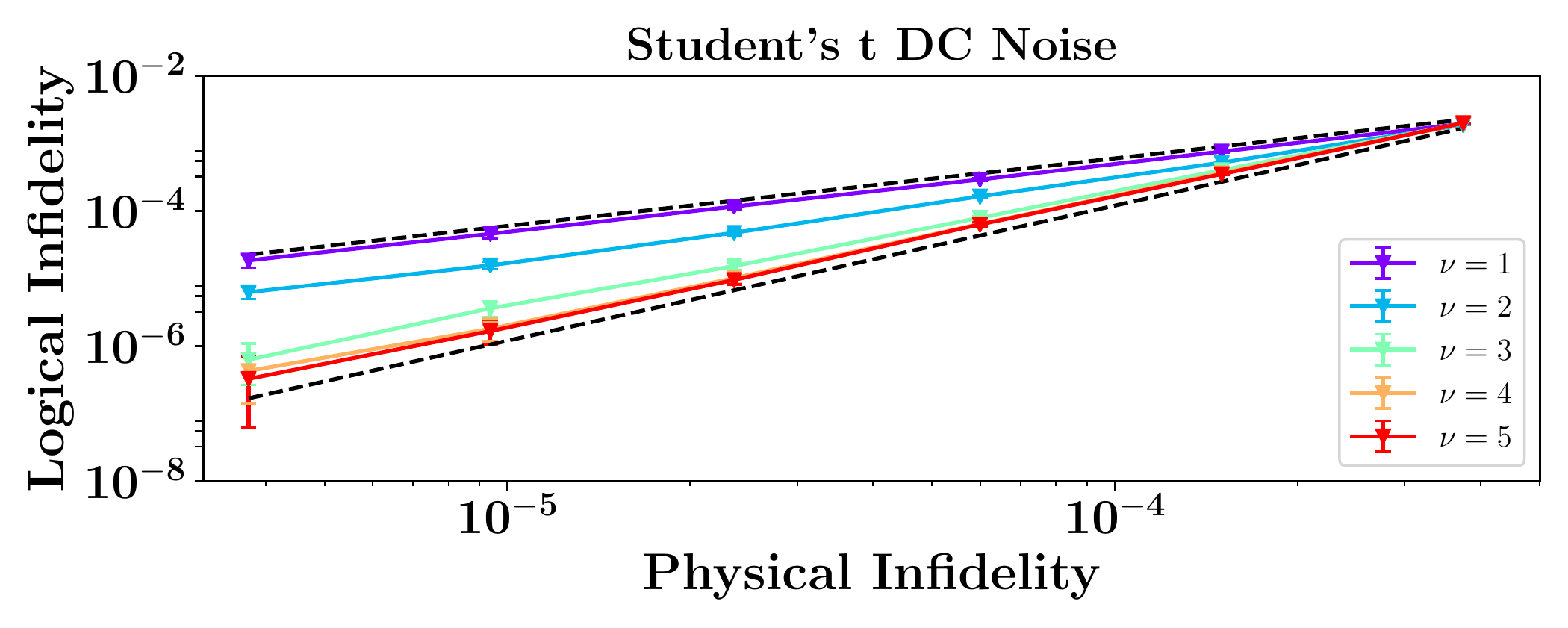}
\caption{\label{fig:student_cheese}Pseudo-threshold plots obtained from simulating the rotated distance-3 surface code with independent (top) and DC correlated (bottom) single-qubit rotation errors drawn from a Student's $t$-distribution for multiple values of the tail index $\nu$ denoted in the legend. For the DC noise case, the curves move from top to bottom in the same order as the legend is displayed with $\nu=1$ having the highest logical error rate and $\nu=5$ the lowest. For white noise, the curves are indistinguishable on this scale. Odd values of $\nu$ correspond to the analytical calculations shown in Table \ref{tab:student_results}. The numerical results agree with our analytical results presented in Tab. \ref{tab:student_results} with the slope of the correlated DC noise logical error rate varying as the tail index increases, while the slope of the uncorrelated white noise is not impacted. Once $\nu \ge 4$ the correlated logical infidelity scales as the square of the physical infidelity in agreement with our analytical predictions. The top and bottom dashed lines show slopes where the logical infidelity is proportional to the physical infidelity and the square of the physical infidelity respectively. Error bars are the 95\% confidence intervals obtained from bootstrap resampling. They are cutoff at the lowest error rates for display purposes.}
\end{figure}

\subsection{Student's t}
Next, we consider the Student's $t$-distribution. We plot the simulation results in Fig. \ref{fig:student_cheese}. We consider the white-noise and DC noise cases and vary the tail index $\nu$ for each simulation. The results are consistent with our analytical predictions. For white (uncorrelated) noise, the code suppresses the logical error rate and is relatively immune to different tail indices. Meanwhile, for DC noise (infinite time correlation), the slope of the logical to physical infidelity depends on the tail index. When $\nu=1$ the distributions have the fattest tails and the simulations show that the logical infidelity is proportional to the physical infidelity. This implies that the \ac{qec} code provides no protection for the encoded qubits. For $\nu=\{2,3\}$ the code offers some protection, but not full protection. Finally, when $\nu \ge 4$ the code offers full protection with the logical infidelity scaling quadratically with the physical infidelity. We note that our analytical results only considered odd values of $\nu$, so the appearance of full code protection was only predicted for $\nu \ge 5$ from our analytical results.

\subsection{L\'evy alpha-stable}
\begin{figure}[t!]
\centering
\includegraphics[width=\columnwidth]{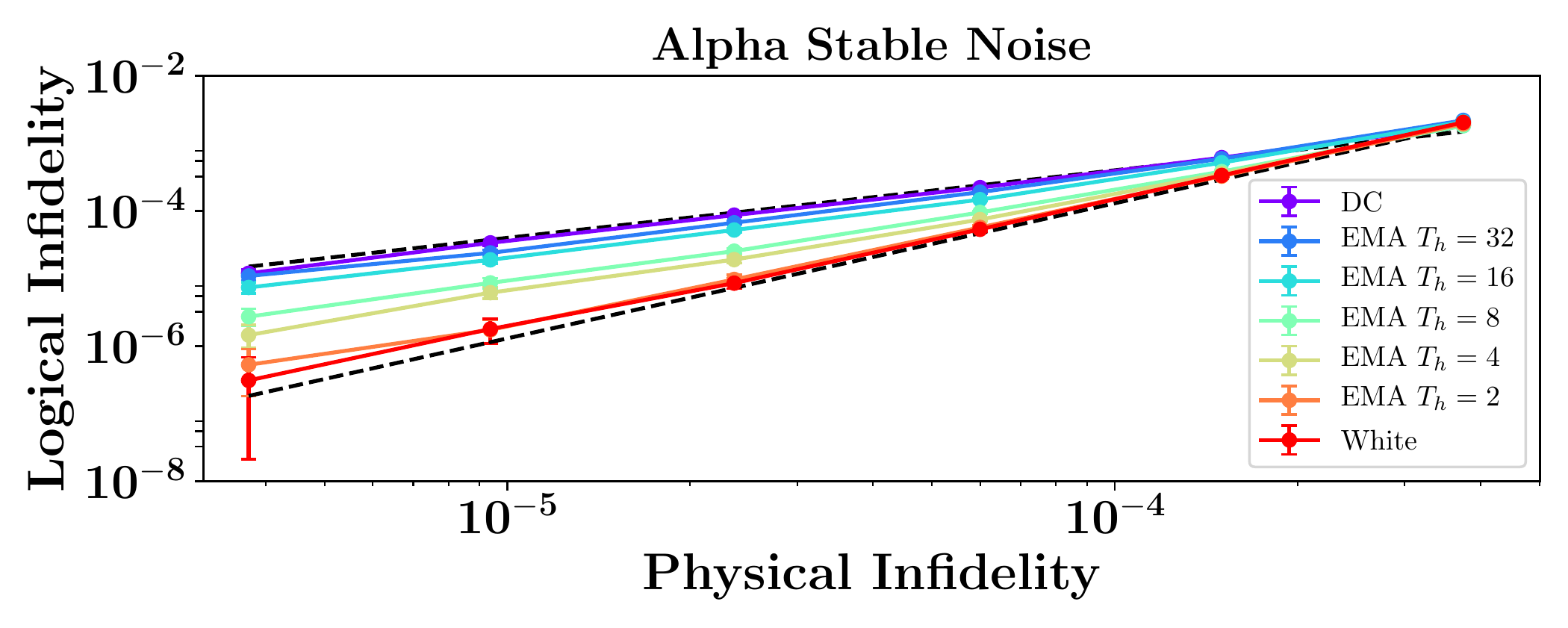}
\caption{\label{fig:stable_cheese}Pseudo-threshold plots obtained from simulating the rotated distance-3 surface code with DC, time-correlated, and independent single-qubit rotation errors drawn from an alpha-stable distribution with $\alpha=1.5$. The curves move from top to bottom in the same order as the legend is displayed with DC noise having the highest logical error rate and White noise the lowest. The numerical results agree with our analytical results with the uncorrelated white-noise case yielding a pseudo-threshold. For DC noise, the code offers no protection with the logical error rate scaling proportionally to the physical error rate. The SchWARMA simulations denoted with titles EMA $T_h$ interpolate between the white and DC noise cases by using an exponential moving average filter. These simulations show that time-correlated noise drawn from a non-Gaussian heavy-tailed distribution can have a strongly detrimental impact to \ac{qec}. The top and bottom dashed lines show slopes where the logical infidelity is proportional to the physical infidelity and the square of the physical infidelity respectively. Error bars are the 95\% confidence intervals obtained from bootstrap resampling.}
\end{figure}

Finally, we show simulations of correlated noise drawn from the L\'evy alpha-stable distribution. Since the distribution is stable, we can use the SchWARMA formalism to simulate time-correlated noise that interpolates between the white noise and DC noise cases. We plot these results in Fig. \ref{fig:stable_cheese}. The results once again agree with the analytical results for the white-noise and DC-noise cases. For intermediate correlations times, enabled by our SchWARMA simulations, we observe that the slope of the logical error rate is related to the half-life of the exponential moving average. The time units of the half life correspond exactly to the gate ticks of the circuit. For the surface code, the syndrome extraction cycles take between 2 and 6 ticks (2-4 CNOT gates for the weight-two and weight-four operators and two rotation gates for the X syndrome), so the slope reduction occurs as the half-life approaches the syndrome cycle as one would expect. The noise begins to generate weight-two errors across syndrome boundaries. We have not examined any mitigation schemes in this paper, but our results suggest that if the noise correlations can be mitigated, through decoupling techniques as one example, that these effects can be reduced and fault-tolerant quantum computing could be realized even if heavy-tailed noise exists.

\section{Conclusions}
We have presented analytical evidence that space- or time-correlated single-qubit noise drawn from heavy-tailed distributions can lead to high-weight errors in a quantum circuit. This can lead to a breakdown of quantum error correction via a reduced code distance (in some instances yielding no protection at all). The exact predictions depend upon the type of distribution used for the noise. 

For Gaussian noise, correlations cause a reduction in leading order coefficient of the pseudo-threshold, but the code can still suppress the noise. This leads to logical error rates scaling as expected with a slope proportional to $\sigma ^{(d+1)/2}$ for a distance $d$ code. Meanwhile for noise with heavy-tailed behavior such as Cauchy, Student's $t$, or L\'evy alpha-stable distributions we find more interesting behavior. The quantum error correcting codes that we considered could not correct correlated noise drawn from Cauchy distribution. This leads to logical failure rates that scale proportionally to the physical error rate. There is no suppression relative to $\sigma$ for any code distance. The Student's $t$-distribution allows us to interpolate between the Gaussian and Cauchy case via the tail-index parameter $r$. As $r$ increases, the error correcting code is able to suppress the logical error rate relative to the physical error rate with increasing power. Once the code distance $d \le 2r-3$ the code achieves its expected error suppression. Finally, the L\'{e}vy alpha-stable distribution also allows us to interpolate between the Gaussian and Cauchy distributions, but since it is a stable distribution we can also consider arbitrary time correlations. Here we find that the correlation time of the noise has a direct impact on the slope of the logical error rate with longer correlation times leading to more reduction in error suppression ability.

These results all reinforce the notion that the ability of quantum error correction to suppress noise is highly dependent on the physical noise model \cite{cafaro2014, Barnes2017, Iyer2018}. It is known that many complex physical systems have non-Gaussian noise and we have pointed to many examples in Sec. \ref{sec:intro} of quantum systems where noise with heavy-tailed statistics is present \cite{Niemann2013,Eliazar2010,Davidsen2002,Kaulakys2005,Lowen1993,Lukovic2008,Kaulakys2013,Ruseckas2010,Kuno2000,Shimizu2001,Pelton2004,Margolin2006,Pelton2007,Mahler2008,Frantsuzov2009,Frantsuzov2013,Haase2004,Schuster2005,Hoogenboom2005,Yeow2006,Lowen1993,Dhariwal1991,Tiedje1981,Lowen1992,PhysPropAmorphMat,Krapf2013,Lobaskin2006,Klein1968, Klein1976, Cizeau1993, Berkov1996, Pickup2009, Neri2010,Sung2019}.  Our results suggest that it is important to develop quantum characterization techniques that can reliably determine the distribution of the underlying noise to ensure the performance of future quantum error corrected systems. Finally, we close with a positive note. Even if heavy-tailed noise with correlations are present in quantum computing systems, our results do not in any way preclude the possibility that quantum control or other physical-level noise mitigation schemes may be able to reduce or eliminate their detrimental impact. We leave that as an another avenue for further research.

\begin{acknowledgments}
Computing resources for the numerical simulations were provided by JHU/APL through the Abyss cluster. We thank Leigh Norris for helpful discussions during the preparation of this manuscript.
\end{acknowledgments}

\end{document}